\documentclass[]{aa}
\usepackage{latexsym}
\usepackage{graphicx}
\usepackage{txfonts}
\usepackage{german}
\begin{document}

\title{Discovery of X-ray flaring on the magnetic Bp-star $\sigma$\,Ori\,E}
\author{D. Groote \and J.H.M.M. Schmitt}
\institute{Hamburger Sternwarte, Gojenbergsweg 112, D-21029 Hamburg, Germany}
\mail{dgroote@hs.uni-hamburg.de, jschmitt@hs.uni-hamburg.de}
\date{Received <date> / Accepted <date>}

\abstract{We report the detection of an X-ray flare on the Bp star
$\sigma$\,Ori\,E with the ROSAT high resolution imager (HRI). 
The flare is shown to have  likely occurred on the
early-type star, rather than on an hypothesized  late-type companion.  
We derive flare parameters such as total energy release, coarse estimates 
of size and density, and also present arguments for a
magnetic origin of the flare.  We place our observations in the
context of a magnetic character of Bp-type stars and speculate on a
common physical basis and connection between Bp and Be stars.}
\maketitle

\section{Introduction}

For cool stars with outer convective envelopes, X-ray emission is
generally considered a proxy indicator of magnetic
activity.   Systematic surveys of X-ray emission
among cool stars in the solar neighborhood have shown the occurrence of
X-ray emission for all stars of spectral type F through M on the main
sequence (Schmitt et al. \cite{Sch:al1}, Schmitt \cite{Sch},
Schmitt \& Liefke \cite{Sch:Lie}).   The observed
X-ray luminosities vary (from star to star) over almost four orders of
magnitude, and the most important parameter governing the X-ray output
level of a given cool star appears to be its rotation rate (Pallavicini et al. 
\cite{Pal:al2}).
The X-ray emission from cool stars is (often) time-variable.  Specifically, 
X-ray flaring is frequently observed among cool stars, and in particular, X-ray 
flaring has been observed for all types of cool stars.  Such X-ray flaring is 
readily explained by the sudden release of magnetic energy and its ultimate 
conversion into heat and radiation.
The X-ray properties of cool stars have to be contrasted to the X-ray properties
of hot stars without outer convective envelopes.  To avoid 
complications from the possibility of X-ray emission
from interacting winds in O-star binary systems, we here restrict our
attention to single stars. Similarly to cool stars, all single hot 
stars appear to be
X-ray emitters, at least down to a spectral type of around B2.  In contrast to
cool stars, the X-ray luminosity of hot stars appears to be independent of 
rotation; rather their X-ray emission is characterized by the relation 
L$_X$/L$_{bol}$ $\approx$10$^{-7}$
(cf., Pallavicini et al. \cite{Pal:al2}).
Thus hot stars can be very strong X-ray sources in terms of their total X-ray 
luminosity L$_X$, however, in comparison to their total radiative output only 
small fractions of their total luminosity are radiated at X-ray energies.

The X-ray emission from hot stars is usually attributed to instabilities
in their radiatively driven winds, leading to stochastic velocity fields and
shocks (cf., Lucy \cite{Luc}).  Surprisingly, despite the stochastic
nature of the X-ray production process in hot stars, their X-ray emission level 
appears to be very stable (cf.,  Bergh"ofer \& Schmitt \cite{Ber:Sc2},  
\cite{Ber:Sc3}) and convincing reports of variable X-ray emission from early type 
stars are rare. For example,
Bergh"ofer \& Schmitt (\cite{Ber:Sc4}) report long-term spectral variability in
the hot supergiant $\zeta$ Ori, which they interpret as a propagating shock 
wave, and Gagn\'e et al. (\cite{Gag:al1}) report a periodic variation of the X-ray 
flux of $\theta ^1$\,Ori\,C, which appears to be modulated with the star's 
rotation period of 15.4 days.  An actual X-ray flare in the B2e star 
$\lambda$~Eri has been reported by Smith et al. (\cite{Smi:al1}); an increase in 
count rate by about a factor of 6 over quiescent values over about 24 hours 
was observed leading to an overall energy release of 2 $\times$ 10$^{36}$ erg.

In general, magnetic fields are thought to be unimportant for the dynamics of 
the outer envelopes of hot stars.   While there are no reasons not to assume the 
presence of magnetic fields on such stars, the typically available magnetic field 
strength upper limits of a few hundred Gauss show that
the magnetic field pressure is much smaller than the ram pressures of their 
winds (Linsky \cite{Lin}). Successful measurements (rather than upper limits)
of magnetic fields in early-type stars are rare. One of the
few examples is $\theta ^1$\,Ori\,C, where a magnetic field of 
about 1.1\,kG has been detected to be modulated with the rotation period of 15.4 
days (Donati et al. \cite{Don:al1}).
The prototypical He-strong He-variable star $\sigma$\,Ori\,E (HD\,37479; spectral 
type B2Vp) is known to have a strong magnetic field of about 10 kG polar field 
strength (Landstreet \& Borra, \cite{Lan:Bor}). A weak stellar wind of some 
$10^{-10}M_{\sun}$\,y$^{-1}$ is channeled along magnetic field lines. In closed
magnetic loops matter is captured, forming torus-shaped clouds (or a ring)
in the plane of the magnetic equator (Groote \& Hunger \cite{Gro:Hu3} (GH1),
Shore \& Brown \cite{Sho:Bro}). These clouds are magnetically coupled to the
star, thus corotate and occult the star twice during one rotational period leading to
absorption in the Stroemgren u-band (Hesser et al., \cite{Hes:al1}),
the higher Balmer lines (Groote \& Hunger, \cite{Gro:Hu1}), and in the
strong resonance UV lines of C\ion{IV}, Si\ion{IV}\
(Smith \& Groote, 2001).  When viewed
face on, the clouds exhibit redshifted emission in H$_{\alpha}$
and some of the above metal lines.
The clouds extend to about 6 $R_{*}$ and are believed to release matter at their
inner boundary back to the stellar surface, thus increasing the He abundance near 
the magnetic equator (Groote \cite{Gro}). First results from 2D MHD
simulations by ud-Doula \& Owocki (\cite{udD:Owo}) support this scenario.

Observations of $\sigma$\,Ori\,E at radio wavelengths (6 cm, Drake et al., 
\cite{Dra:al1}) are consistent with the assumption of synchrotron radiation by 
mildly relativistic electrons. X-rays from $\sigma$\,Ori\,E have been observed by 
Bergh"ofer \& Schmitt (\cite{Ber:Sc1}) at a constant level of about $8.2\times 
10^{-3}$ counts\,sec$^{-1}$ using the ROSAT high resolution imager (HRI).  The 
origin of this X-ray emission is unclear.

Converting the observed X-ray count rate to an X-ray flux yields a 
spectrum-dependent estimate of approximately 
$2\times 10^{-13}$ erg\,cm$^{-2}$\,sec$^{-1}$ 
(appropriate for $T\approx 20$\,MK and 
$N_{\rm H}\approx 10^{21}$\,cm$^{-2}$), which corresponds to a fraction
$\approx 3.9\ 10^{-7}$ of 
the total bolometric flux of $\sigma$\,Ori\,E.  $\sigma$\,Ori\,E thus lies in 
line with the typical values observed from O-type stars.  Alternatively, the
magnetic geometry of $\sigma$\,Ori\,E might be relevant and the X-ray emission 
might originate from frictional heating in the wind when helium and hydrogen decouple
from the driving metals (Krti\v{c}ka J.\& Kub\'at, \cite{Kri:Kub}) and/or
from the impact of wind particles into the ring/clouds (Babel \& Montmerle,
\cite{Bab:Mon}). Since the above described ring is constantly filled with
new material from the star, matter also has to be released and it is natural to
assume magnetic field line reconnection with heating to some 10$^{6}$~K in
analogy to the Earth's magnetosphere. This release of matter is not expected
to occur continuously, but rather sporadically,  leading to variable
X-ray emission.

However, in contrast to these expectations, first observations (duration about 
15 ksec) with the ROSAT HRI (Bergh"ofer \& Schmitt \cite{Ber:Sc2}, 
\cite{Ber:Sc3}) did not show any significant temporal variations in the X-ray 
flux from $\sigma$\,Ori\,E. Since then more data on $\sigma$\,Ori\,E 
(about 75 ksec distributed over more than 30 days) were taken with the ROSAT 
HRI within the context of another program and it is the purpose of this paper 
to present (Sect.\,2) and discuss these observations (Sect.\,3). We determine
basic flare properties and will in
particular compare the $\sigma$\,Ori\,E data
with the flare found in $\lambda$~Eri (Smith et al.
\cite{Smi:al1}).  Our conclusions will be drawn in Sect.\,4, and finally
we speculate on a possible connection between Bp and Be stars.

\section{Observations and data analysis}

A series of pointings with the ROSAT HRI was carried
out on the multiple O-type system  $\sigma$\,Ori\,AB. The ROSAT HRI detector
has very little intrinsic energy resolution, but very good angular
resolution ($\approx$ 5 arcsec).  A separation of the X-ray emission
from $\sigma$\,Ori\,AB and $\sigma$\,Ori\,E was therefore
straightforward, but no information
on the shapes of the X-ray energy distributions of $\sigma$\,Ori\,AB and
$\sigma$\,Ori\,E was available. The ROSAT
observations were carried out as a sequence of individual snapshots,
each lasting approximately 2000 seconds, separated by approximately
24 hours.  The total time span of the X-ray observations covered 30 days from
JD 2449775 - JD 2449808, and the total source integration time was about
74 ksec.

The data analysis was carried out within the EXSAS content in the MIDAS
environment; the ROSAT sequences analyzed were the ROR numbers 
201882 - 201915.  The photon
event tables and event rates files from the sequences were merged
with the EXSAS command MERGE into one effective observation and an image
containing the data in the pulse height channels \discretionary{}{}{2-9} was 
created. From this image the photons originating from $\sigma$\,Ori\,E and
$\sigma$\,Ori\,AB as well as a background region were extracted; for the
light curves we used a circular extraction region of 13 arcsec, while for
the background a much larger region in the central part of the detector was
chosen.  From the source photon files light curves were generated with
corrections for dead time and vignetting; note that these latter corrections
are very small and are far below the statistical errors in each
time bin.

\begin{figure}
\centering
\includegraphics[height=10.5cm]{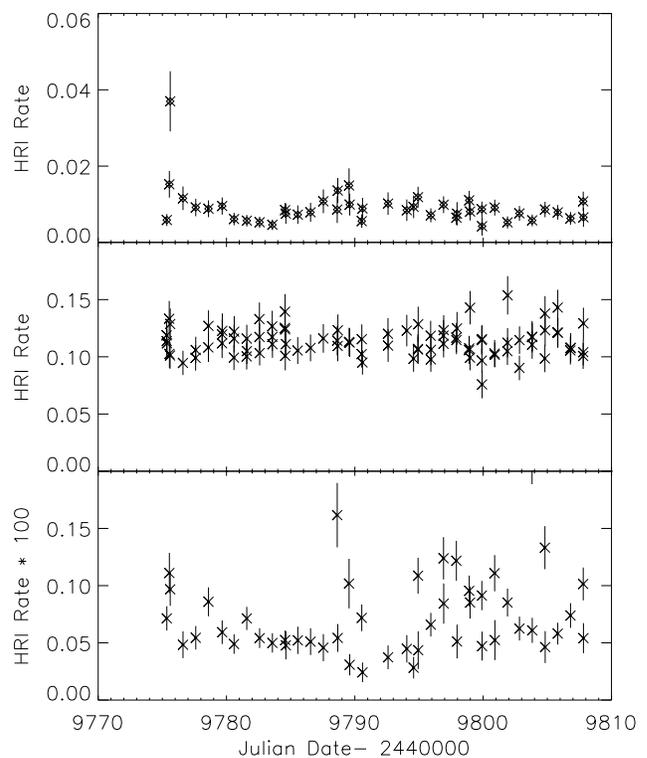}
\caption{ROSAT HRI light curves of $\sigma$ Ori E (top panel), $\sigma$ Ori AB 
(medium panel), and a representative background region (bottom panel); note that 
the background rate has been increased by a factor of 100 for better visibility. 
Note also that background contributions to the source count rates are very small.}
\label{fig1}
\end{figure}

\subsection{X-ray light curves}

\begin{figure}
\includegraphics[height=6.3cm]{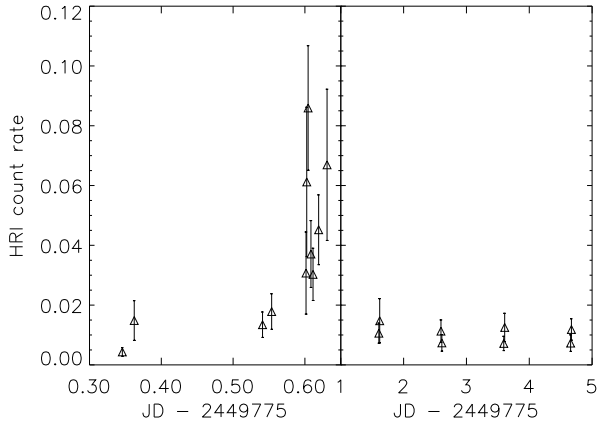}
\caption{ROSAT HRI light curve of $\sigma$ Ori E during flare increase (left 
panel)and constant X-ray fluxes on following observations (right panel).
Note the different time scales and the jump between the two panels.}
\label{fig2}
\end{figure}

The generated, but not background subtracted, light curves for
$\sigma$\,Ori\,E, $\sigma$\,Ori\,AB and the background are shown in Fig. 1
as count rate versus modified Julian date. Note that the background count rate was very
low, about a factor 100 below that of $\sigma$\,Ori\,E and about a
factor 400 below that of $\sigma$\,Ori\,AB; although the background count rate
does vary, its variations do not influence the source light curves.
As is clear from Fig. 1, the light
curve for $\sigma$\,Ori\,AB stays constant, while in the case of 
$\sigma$\,Ori\,E the first data points are low and comparable to the X-ray 
emission level observed by Bergh"ofer et al. (\cite{Ber:Sc1}).
Then the count rate increases rapidly over the next data bins, and
returns to the pre-flare level with possibly some small variations.
In Fig. 2 we show an enlargement of the X-ray light curve during Julian day
2449775. The increase in X-ray count rate took place on a time scale of at most 
2.5 hrs.  Whether our observations cover the peak of the flaring emission is 
unclear; 24 hours later, at JD = 2440776.55, when the 
next ROSAT observations were taken, the emission was back at pre-flare levels.  
It is very instructive to over-plot the ROSAT X-ray light curve of 
$\sigma$\,Ori\,E with the ROSAT PSPC light curve of $\lambda$ Eri reported by 
Smith et al. (\cite{Smi:al1}; diamonds in Fig. \ref{fig2}); the
$\lambda$ Eri (PSPC) data were appropriately shifted in time and scaled to the
same apparent pre- and post flare levels.  As is obvious from Fig. \ref{fig2}, 
the $\sigma$\,Ori\,E flare light curve is consistent with the observed light 
curve morphology of $\lambda$ Eri.  

The statistical significance of the observed count rate increase in 
$\sigma$\,Ori\,E is extremely high.  If we assume a quiescent rate of 0.008 (HRI) 
cts\,sec$^{-1}$ (which may actually be a little higher than observed), we expect 
23.1 photons in the 2886.5 sec time interval between JD 2449775.563 and JD 
2449775.631. This has to be contrasted with the total number of 82 photons,
that were recorded during that time interval.  Assuming Poisson statistics,
the probability to observe 82 photons with 23.1 expected is almost zero.
In Fig. 2 one recognizes a more or less linear increase with
two data points (at JD 2449775.65) above the linear curve.  We analyzed 
 - using a Kolmogorov-Smirnov test -
the individual photon arrival times during this short data interval and 
found that deviations from constancy are possible at the $\approx$ \,95\% 
confidence limit.
Thus shorter time scale variability is possible but not necessarily required
by our HRI observations of $\sigma$\,Ori\,E.

\begin{figure}
\includegraphics[height=6.3cm]{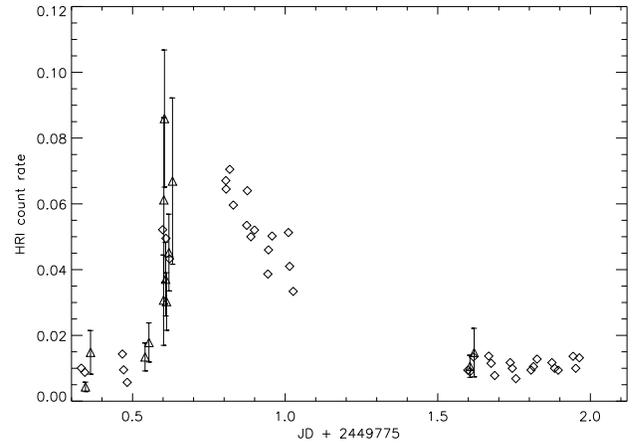}
\caption{Superposition of ROSAT HRI light curve of $\sigma$\,Ori\,E (triangles, cf., 
\ref{fig2}) and the ROSAT PSPC light curve of $\lambda$\,Eri (open diamonds, cf., Smith et al. 
\cite{Smi:al1}).  Note the morphological similarity between the two light 
curves in the rise phase of the flare.}\label{fig3}
\end{figure}

\subsection{Phasing of the X-ray emission and of the X-ray flare}

\begin{figure}
\includegraphics[width=8.8cm]{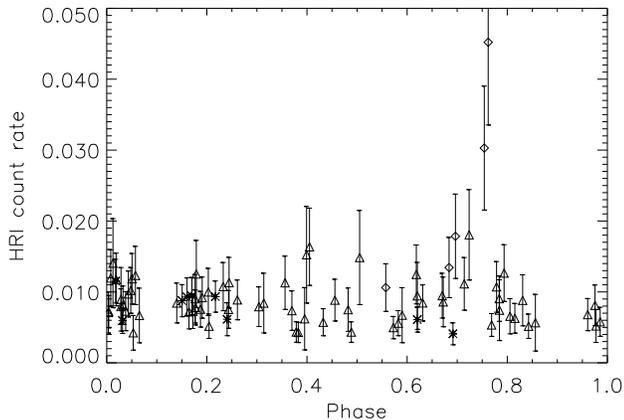}
\caption{Phased ROSAT HRI light curve of $\sigma$ Ori E; note that the high data 
points(diamonds) are due to the flare shown in Fig. \ref{fig2}. While in new 
(triangles) and old data points (asterisks) no phase dependence of the X-ray 
emission is detectable.}
\label{fig4}
\end{figure}

The rotation period of $\sigma$\,Ori\,E is well known. Using
the ephemeris $E = 24442778.819 + n \times 1.190815$ days, we can assign a phase 
to each measured data point.  The choice of phase is such that $\phi = 0$
corresponds to the phase with maximum absorption in the $u$-band (Hesser et 
al.\cite{Hes:al1}) which corresponds to the magnetic null phase 
(Landstreet \& Borra,\cite{Lan:Bor}).
The corresponding plot is shown in Fig. \ref{fig4}.  We have distinguished
between flare data points (diamonds) and other data (triangles). Data points 
taken from an earlier HRI observation of $\sigma$\,Ori\,E in September 1992, 
that was already used by Bergh"ofer \& Schmitt (\cite{Ber:Sc1}), are represented 
by asterisks.
As is clear from Fig. \ref{fig4}, except for the flare described earlier, no 
rotational variability of $\sigma$\,Ori\,E can be recognized.  This is in marked 
contrast to the case of the magnetic early type star $\theta^1$Ori\,C, whose 
X-ray emission is modulated with the rotational period of 15.4 days (Gagn\'e et al. 
\cite{Gag:al1}). 
Donati et al. (\cite{Don:al1}) explain this variability of $\theta^1$\,Ori\,C in 
the framework of the magnetic confined wind-shock (MCWS) model of Babel \& Montmerle
(\cite{Bab:Mon}, \cite{Bab:Mo2}) where the observed 
modulation is essentially due to the eclipse of the post-shock region by the star
itself. No such modulation is apparent in the phased X-ray light curve of $\sigma$\,Ori\,E.

\subsection{Energetics of the X-ray flare}

Clearly, the ROSAT HRI observations did not fully cover the
flare on $\sigma$\,Ori\,E, and therefore
the energetics of this X-ray flare cannot be directly measured.  The
observed X-ray light curve can be approximated by a straight line, linearly 
increasing from 0.008 cts\,sec$^{-1}$ to 0.06 cts\,sec$^{-1}$ over a time 
span of 8640 seconds (i.e., 0.1 days).
We specifically note that the flare observations consist of three time 
intervals: During a first time interval lasting 1163 sec 15 photons were recorded 
(the 2 light curve data points in Fig. 2 at 0.54 days); during a second 
time interval yielding only 105 seconds of exposure 3 photons were recorded 
but the bulk of the flare increase occurred in the third time interval (after 
0.60 days in \ref{fig2}) with a duration of 
1490.5 seconds, during which 69 photons were detected (i.e., the data points 
between 0.60 and 0.62 in Fig. 2); a fourth data interval of 120.5 seconds yielded 
8 photons (i.e., the large error bar data point at 0.63 days in Fig. 2).  The 
next data were taken one day later with the count rate being back to  
pre-flare level. Had this whole increase been continuously observed by ROSAT, a 
total of 225 counts would  have resulted from the flaring X-ray emission 
(assuming a linear increase in X-ray count rate). The 
observed flare peak was about 0.052 cts\,sec$^{-1}$, which of course need not 
be the actual flare peak, and the whole flare decay remained unobserved by ROSAT. 
Thus, all we can state is that the flare decay time was 24 hours at most. 
Assuming then that the count rate started to
decrease linearly immediately after the observed flare peak to the
pre-flare level 24 hours later, we compute a total
of 2250 counts attributable to the flare decay and hence a total of 2475 counts 
to the whole flare.  Alternatively, if we assume that flare rise and decay have 
the same number of counts, the whole flare would have had 450 HRI counts.   With 
a distance of 640 pc towards $\sigma$\,Ori\,E (Hunger et al., \cite{Hun:al1}) and 
a flux conversion of 2.4 $\times$ 10$^{-11}$ erg\,cm$^{-2}$\,count$^{-1}$, 
which is appropriate for one 
recorded HRI count corresponds to 1.2 $\times 10^{33}$ erg and hence the total 
flare energy $E_{\rm tot}$ emitted at X-ray wavelengths should be 
5.3 $\times$ 10$^{35}$ erg 
and 29 $\times$ 10$^{35}$ erg for the two scenarios discussed.  These numbers are 
clearly not mathematically correct lower and upper bounds for the total flare, 
however, we think that the true value is very likely bracketed by these numbers. 
This view is supported by an analysis of the $\sigma$\,Ori\,E flare reported
by Pallavicini et al. (\cite{Pal:al1}); from their $\sigma$\,Ori\,E light curve
we estimate a total X-ray energy release of $\approx 1.3 \times 10^{36}$ erg  
using a distance of 640 pc and a (XMM-Newton) flux conversion factor of 
5.4 $\times$ 10$^{-12}$ erg\,cm$^{-2}$\,count$^{-1}$ (Schmitt et al. \cite{Sch:al3}).
These numbers must be lower bounds to the actual flare energetics, 
since thermal, kinetic and magnetic energy are not included. Although there are 
flares on cool stars with even larger energies emitted (more than 10$^{37}$\,erg 
for a long-duration flare on the RS CVn binary CF Tuc (K"urster \& Schmitt 
\cite{Kue:Sch}) and the Algol-type binary $\beta $ Per (Schmitt \& Favata 
\cite{Sch:Fav})), the above considerations show that
the flare on $\sigma$\,Ori\,E is still among the most energetic
stellar X-ray flares hitherto observed. 
The observed peak X-ray luminosity was 6.1 $\times$  10$^{31}$\,erg\,sec$^{-1}$,
again reaching values observed in the strongest cool star X-ray 
sources.  

\section{Interpretation and discussion}

\subsection{Unseen companions ?}

Pallavicini et al. (\cite{Pal:al1}) attribute a flare observed with XMM-Newton apparently
from $\sigma$\,Ori\,E to an unseen companion. While the stellar cluster around
$\sigma$ Ori\,AB is indeed very young and young stars are known to be capable
of producing strong X-ray flares, $\sigma$\,Ori\,E most
probably does not belong to the $\sigma$\,Ori cluster. Hunger et al. (\cite{Hun:al1})
derived fundamental parameters $T_{\rm eff}$ and log~$g$ for the stars 
$\sigma$\,Ori\,E and $\sigma$\,Ori\,D, using high resolution CASPEC data.
The latter star turned out to lie
near the zero age main sequence (log~$g=4.3$) with a spectroscopic distance
$d=370$~pc, which is in excellent agreement with the Hipparcos distance of 
$d = 352$\,pc for $\sigma$\,Ori\,AB,
while for $\sigma$\,Ori\,E log~$g=3.95$ and a distance of $d = 640$\,pc was reported,
clearly identifying $\sigma$\,Ori\,E as a background star. Using evolutionary tracks
from Schaller et al. (\cite{Sch:al0}) with $T_{\rm eff}=22500$~K and
log~$g=3.95$, $\sigma$\,Ori\,E is a 9~M$_{\sun}$ star with an age of 17~Myr.
Thus both, distance and age make a membership to the $\sigma$\,Ori cluster unlikely.

The angular resolution of our HRI data is much better compared to the XMM-data 
and the relative X-ray positions of $\sigma$\,Ori\,AB and E agree very well with 
their relative optical positions ($\rho_{\rm opt}= 41.36 $\,arcsec, 
$\rho_{\rm X}= 41.07 \pm 0.27$\,arcsec, and position angle 
$\alpha_{\rm opt} = 61.55$\,deg, $\alpha_{\rm X} = 61.60 \pm 0.27$\,deg). 
Because of this excellent agreement of X-ray and optical positions, we 
consider a chance alignment of  $\sigma$\,Ori\,E with a low mass member of 
the closer $\sigma$\,Ori cluster unlikely.

A positional agreement of only 0.3 arcsec translates into a projected distance
of 300\,au of a hypothesized companion of $\sigma$\,Ori\,E.
Thus $\sigma$\,Ori\,E could be a spectroscopic binary itself.  
However, in the high-resolution, high-signal-to-noise FEROS spectra 
(SNR$\approx$250) of
$\sigma$\,Ori\,E (c.f. Reiners et al. \cite{Rei:al1}) no sign of any 
companion can be found. We estimate that any G-star hidden in the spectrum 
must be smaller than about $0.2\,R_{\sun}$ with correspondingly larger limits 
for M-type stars. If an hypothesized companion of $\sigma$\,Ori\,E is indeed 
very young (say 2 Myrs), its radius should be far larger and therefore be 
visible in the spectrum, contrary to what is observed.
Further, radial velocity measurements by Groote \& Hunger (\cite{Gro:Hu2}) 
show that any RV velocity variations would have to stay below 1 km\,s$^{-1}$. 
And finally, absolute flux measurements from UV to IR, taken from a region 
encompassing a substantial volume around $\sigma$\,Ori\,E, agree very well 
with the fluxes calculated from appropriate model atmospheres and provide 
no evidence whatsoever for a hidden companion or a foreground star possibly 
belonging to the $\sigma$\,Ori system. Nevertheless, a, say, $\le 0.5 M_{\sun}$ 
main sequence (i.e. ``old'') star at $\approx$300\,au might be hidden in the 
spectroscopic and photometric data, and could thus 
be held responsible for the  observed X-ray emission. Such a star, however, 
would have an age of $\approx$\,20\,Myr rather than 2\,Myr, 
and hence be comparable to clusters like IC\,2602 (Randich et al.\cite{Ran:al1}) 
or IC\,2391 (Patten \& Simon \cite{Pat:Sim}). G and K-type stars in 
those clusters show X-ray luminosities about a decade lower than observed for 
$\sigma$\,Ori\,E. Further, the X-ray flux of $\sigma$\,Ori\,E is quite constant 
outside  the flare, which is rather atypical for a very young late type star. 
Also, the observed flare energetics are unusual (albeit not unthinkable) for 
a 20\,Myr late type star. Attributing the observed quiescent and flaring 
X-ray emission from $\sigma$\,Ori\,E to a physical companion would thus imply a 
rather unusual companion star; note that $\sigma$\,Ori\,E is already quite 
unusual by itself. Therefore, we consider this possibility again unlikely.
For the following we therefore assume instead that the observed X-rays and 
flares are produced in the circumstellar environment of $\sigma$\,Ori\,E.

\subsection{Wind-shock X-ray emission} 

It seems reasonable to compare $\sigma$\,Ori\,E to other similar stars. As far as 
the quiescent flux of $\sigma$\,Ori\,E is concerned, it is comparable to that 
observed from other early type stars.  However, $\sigma$\,Ori\,E is a chemically 
peculiar star with a magnetic dipole field of about 10 kG polar field strength.  
X-ray emission has been observed from two other magnetic
stars, i.e.,  $\theta^1$\,Ori\,C, which is much hotter ($T_{\rm eff}\approx 
45\,000\,$K), and IQ Aur which is much cooler ($T_{\rm eff}\approx 15\,000\,$K) 
than $\sigma$\,Ori\,E ($T_{\rm eff} = 23\,500\,$K).   Their X-ray production 
can be successfully explained by the MCWS model (Babel \& Montmerle \cite{Bab:Mo2},
Donati et al. \cite{Don:al1}); in particular, this model provides a 
natural explanation for the observed periodicity in the X-ray flux of
$\theta^1$\,Ori\,C. 
X-ray flaring for B-type stars  has been reported only for $\lambda$\,Eri 
(spectral type B2e; Smith et al. \cite{Smi:al1}), however, $\lambda$\,Eri is so 
far not known to be a magnetic star.

The central feature of the MCWS model of Babel \& Montmerle is the
channeling of the stellar wind emanating at high magnetic latitudes by the 
magnetic field.  The magnetic field forces the high-latitude wind into the 
magnetic equator plane, where the the wind streams from both hemisphere collide.  
The shocked gas leads to the observed X-ray emission in the post-shock region; 
the gas eventually cools and forms a thin cooling disk in the magnetic equator 
plane.  This model would appear to provide an adequate description for 
$\sigma$\,Ori\,E at first sight, too, however, the constant X-ray flux 
of $\sigma$\,Ori\,E (cf, \ref{fig4}) has to be contrasted with the periodic variation 
of $\theta^1$\,Ori\,C.  
In the latter case the phase variation is interpreted 
by Donati et al. (\cite{Don:al1}) as a partial eclipse of the X-ray emitting 
post-shock region near to the stellar surface at a distance of about $1.5\,R_*$.
The magnetic field of 
$\theta^1$\,Ori\,C is much smaller ($\approx 1.1$ kG) than that of 
$\sigma$\,Ori\,E ($\approx 10$ kG), while, on the other hand, the wind 
from $\theta^1$\,Ori\,C is much stronger due to the intense O star radiation 
field. Consequently, one would expect a much stronger magnetic confinement in the 
case of $\sigma$\,Ori\,E, the cooling disk should be denser and extend further 
out (up to 6 $R_*$ as observed). Therefore, X-ray absorption at magnetic pole 
phases with the disk viewed face-on might be expected. 
The magnetic geometry of $\sigma$\,Ori\,E ($i = 54$ deg, 
$\beta = 67$ deg) implies that twice during each rotation cycle the disk is 
viewed edge-on, when X-ray emission should be at maximum, while at least at one 
phase ($\Phi = 0.73$) the disk is viewed face-on, where X-ray emission should be 
at a minimum. Such a variation is clearly not observed (cf., \ref{fig4}), and 
the model successfully explaining 
the phase variation of the X-ray flux of $\theta^1$\,Ori\,C does not appear to be 
applicable for the case of $\sigma$\,Ori\,E.

What is the reason for the failure of this model describing well the
X-ray production in a hotter as well as in a cooler magnetic star? The mass
loss of the three stars considered, i.e., IQ Aur, $\sigma$\,Ori\,E, and
$\theta^1$\,Ori\,C increases from about 10$^{-11} M_{\sun}$\,y$^{-1}$ over some 
10$^{-10}M_{\sun}$\,y$^{-1}$ to about some 10$^{-7}\,M_{\sun}$\,y$^{-1}$, 
respectively. The low mass loss rate for IQ Aur is the result of non-solar 
abundances in the wind.  Helium (because it is partly neutral) will decouple 
already in the photosphere, and thus never leaves the stellar surface. Hydrogen 
decouples near the stellar surface before reaching escape velocity, and will 
thus be re-accreted if its velocity is too small to reach the magnetic equator. 
The He-depletion at the magnetic pole (Babel \&Montmerle \cite{Bab:Mon}) seems 
more likely to be an H-enrichment due to the re-accretion of hydrogen falling 
back to the surface (see Groote \cite{Gro}). The metals instead receive sufficient 
momentum from the radiation field to be accelerated against
gravity to velocities of more than $\approx 800$ km\,sec$^{-1}$, and have 
therefore sufficient kinetic energy to produce the observed X-ray emission.
In the case of the O star wind of $\theta^1$\,Ori\,C, radiation pressure on the 
driving metals is sufficient to accelerate both metals and the passive plasma, 
which is coupled to the metals through Coulomb forces, to nearly twice the 
escape velocity ($\approx 2500$ km\,sec$^{-1}$). In the transition region between 
massive O star winds, where metals, hydrogen, and helium are fully coupled, and the 
pure metallic winds of the cool B stars with essentially no coupling between 
metals and hydrogen and helium, the mass loss rate decreases with decreasing 
effective temperature and decreasing wind velocities.  Hunger \& Groote 
(\cite{Gro:Hu4}) first proposed that the He-enrichment in the atmospheres
of He-variable stars may be the result of He-decoupling in the winds of these 
stars.  After decoupling, helium will be re-accreted to form He-rich spots on the 
stellar surface. Later they extended the idea to He-weak stars (Hunger \& Groote,
\cite{Hun:Gro}, see also Groote, \cite{Gro}), where hydrogen is also decoupled in the 
wind and may be re-accreted at the magnetic poles. Such a decoupling of passive 
elements seems inevitable if wind velocities drop below escape velocity. The 
coupling becomes weaker with the decrease of density with increasing distance from 
the star. Once helium (or hydrogen) is decoupled, the remaining wind particles may
be efficiently accelerated further because of the substantial decrease in mass left for
acceleration.
In the case of $\sigma$\,Ori\,E the initial wind velocities are probably low, 
additionally depending on magnetic latitude, and are perhaps of the order of only 
$200 - 300$\,km\,sec$^{-1}$. Therefore, heating in a post-shock region may be 
insufficient to produce any significant X-ray emission, so that other 
heating processes are required. Since no phase variation is observed 
(Bergh"ofer \& Schmitt, \cite{Ber:Sc1}, and this work),
we conclude that the source of the observed X-rays is not occulted either by the star
nor by the disk, 
and must therefore be located outside the closed loops of the magnetosphere. 
This assumption also fits the finding that wind absorption seen in the 
resonance lines of \ion{C}{iv} and \ion{Si}{iv} does not vary with phase 
either (Groote \& Hunger \cite{Gro:Hu4}, Smith \& Groote \cite{Smi:Gro}).
Groote \& Hunger considered an expanding corona outside a wind-fed 
magnetosphere, which releases matter more or less regularly from the corotating 
clouds.

\subsection{Size and location of the flaring region}

In view of the above considerations it is likely that also the site of the 
X-ray flare is located rather far away from the stellar surface.  Unfortunately, 
the essential light curve parameters such as peak flux and decay time are only 
rather indirectly available from the ROSAT data (see above), and the flare 
temperature is unknown.  For the following we adopt an {\it ad hoc}
temperature of 10$^7$\,K, which is a typical temperature for cool star flares. 
Fortunately, the cooling function of optically thin plasma is rather insensitive 
to the precise value of temperature in that range.   If we further adopt the 
observed peak X-ray flux of 6.1 $\times$ 10$^{31}$ erg\,sec$^{-1}$ as the actual 
peak flux,  we compute a minimal emission measure of $EM_{\rm min} \approx 
10^{54}$\,cm$^{-3}$.   As to the decay time,
$\tau_{\rm decay}$, it could be as long as $\tau_{\rm decay} \approx $ 40 ksec 
if indeed the same morphology as for the flare on $\lambda$\,Eri applies, but 
also much shorter. If we adopt $\tau_{\rm decay} = $ 40\,ksec and further assume 
that the flare cools only through radiative cooling we can compute a characteristic 
electron density $N_{\rm e}$ from the relation
\begin{equation}
N_{\rm e} = \frac {3 k T} {P(T) \tau_{\rm decay,}}
\end{equation}
where $k$ denotes Boltzmann's constant and $P(T)$ the plasma cooling function.
For $T \approx$ 10$^7$\,K, we find $P(T)$ $\approx$ 10$^{-23}$ 
\,erg\,cm$^3$\,sec$^{-1}$and hence $N_{\rm e} \approx 10^{10}$\,cm$^{-3}$. We 
consider this a minimal plasma density, since both higher temperatures (which are 
usually observed in cool star flares) and shorter decay times (40 ksec is close 
to the maximally possible) will lead to higher densities. Some questions remain 
about the correct cooling function $P(T)$; the values 
used refer to a fully ionized plasma with cosmic abundances in collisional 
equilibrium, and these assumptions may not fully apply. Adopting then $N_{\rm e} 
= 10^{10}$\,cm$^{-3}$ and  $EM_{\rm min} = 10^{54}$\,cm$^{-3}$, we compute the 
flaring volume $V_{flare} = 10^{34}$\,cm$^{3}$, which corresponds to a 
sphere with radius $1.3\times 10^{11}$\,cm or $0.4\,R_{*}$.  Thus the flaring region 
is indeed small compared to the star itself, and its presumed distance from the 
stellar surface. We emphasize that other cooling mechanisms will imply even 
larger densities and correspondingly smaller volumes.  Assuming a field scale 
size of $1 R_{*}$, the magnetic field strength at $6 R_{*}$ will be about 50\,G 
yielding a pressure of somewhat below 100\,dyn\,cm$^{-2}$; the thermal pressure 
of plasma with the above assumed temperatures and densities is about 
30\,dyn\,cm$^{-2}$, i.e., of the same order.   Calculating the speed $v_{\rm 
wind}$, that a plasma with density of $N_{\rm e} = 10^{10}$\,cm$^{-3}$ must have 
to produce a ram pressure of the same magnitude, results in a value of  $v_{\rm 
wind} \approx 10^8$\,cm\,sec$^{-1}$, which would be a plausible wind speed.

Assuming a plasma ``cloud'' 6 $R_{*}$ away from the surface with a thermal 
pressure of 30 dyn\,cm$^{-2}$ is not unreasonable.  Groote \& Hunger (\cite{Gro:Hu3}) 
derive an electron densitiy of $\approx 10^{12}$\,cm$^{-3}$ at 
temperatures $\approx 10^{4}$\,K, yielding
thermal pressures of the same order as the hot X-ray emitting gas. Also, the 
assumption of magnetic reconnection is very plausible. Adopting the above derived
values for $V_{\rm flare}$ and $E_{\rm tot}$ we can compute the total 
reconnected field component $B_{\rm recon}$, which is found to be 60\,G,
i.e., again of the same order as the (vacuum) dipole field. Thus, a
magnetic reconnection scenario requires field strengths and pressures 
which are entirely plausible.

\subsection{Heating}

The most commonly adopted heating process for single O and B stars are 
shocks formed by instabilities in the radiatively driven winds of these stars.
Porter \& Skouza (\cite{Por:Sk1}) used 1D hydrodynamic calculations to
model the case of $\sigma$\,Ori\,E. According to Porter \& Skouza 
(\cite{Por:Sk1}) hydrogen decouples to form  shell-like structures which are 
then shocked and become the source of the observed X-rays. 
Krti\v{c}ka \& Kub\'at (\cite{Kri:Kub}) used 2D hydrodynamic calculations 
(albeit not fully self-consistent, and neglecting magnetic fields and rotation), 
and found that decoupling of helium is possible in B3 stars when modeling the 
outflow as a four component gas. Similarly, their models showed decoupling in 
B5 stars in the context of a 3 component gas. In both cases heating processes 
are important for the remaining wind component, and temperatures of several 
$10^6$\,K may be found at larger distances. In the case of magnetic stars the 
wind will be slowed down additionally by the magnetic latitude dependent 
channeling of the magnetic field, and their results may also be valid for 
stars with somewhat higher temperatures. For the case of $\sigma$\,Ori\,E 
the heating of the ionic component would then occur outside the closed magnetic 
field lines in an approximately radial magnetic field line configuration, forming 
a corona as already assumed by Havnes \& Goertz (\cite{Hav:Goe}) and GH1. 
In that case only a small part of the spherical X-ray emitting region with 
a diameter in excess of $6 R_*$ will be occulted by the dense gas ring. 
This view is also supported by VLBI observations of $\sigma$\,Ori\,E 
presented by Phillips \& Lestrade (\cite{Phi:Les}), who find a scale size of 
6 to 10 $R_*$ for the radio emitting regions, once the new (larger) 
distance towards $\sigma$\,Ori\,E is taken into account (see also Hunger et al. 
\cite{Hun:al2}). Such heating processes will not be present in 
a star like $\theta^1$\,Ori\,C whose wind is dense and whose wind components 
are all well coupled. Within the context of such a scenario a 
plausible explanation for the observed X-ray flare(s) can thus be 
found. 
The permanently ongoing filling of the ring/clouds by the wind will increase 
the plasma ram pressure, while the magnetic pressure stays constant. Once the 
ram pressure exceeds the magnetic pressure, matter cannot remain confined by the
field, and must be released from the magnetosphere.  The reconnection of magnetic 
field lines occurring during this release then provides the additional heating 
with ensuing X-ray emission. GH1 estimated the time required to refill the 
clouds using the mass loss and found a time scale of the order of about a month. 

\section{Summary and outlook}

The discovery of a X-ray flaring in the Bp-type star $\sigma$\,Ori\,E
with a well measured magnetic field of 10 kG provides strong evidence for 
magnetic field related heating in the magnetosphere of $\sigma$\,Ori\,E.  
The fact that yet another flare on $\sigma$\,Ori\,E has been observed with 
XMM-Newton (Pallavicini \cite{Pal:al1}) indicates that such flares are not that 
rare and may occur more or less regularly at least on $\sigma$\,Ori\,E.  
The other early type star with a reported flare is the Be-type star 
$\lambda$\,Eri.  Given the similarity of stellar parameters of 
$\sigma$\,Ori\,E and $\lambda$\,Eri and their winds, one wonders about 
possible similarities in the physical processes encountered in both types of 
stars. 

Is there then a connection between Bp and Be stars? Magnetic fields have 
already been introduced as a possible source for some of 
the observed features of Be-type stars and the formation of disks in presence 
of a magnetic field has been analyzed by Cassinelli et al. (\cite{Cas:al1}) 
and Maheswaran (\cite{Mah}). We argue here that magnetic fields 
actually provide the physical link between the two classes of objects. 
The interaction of wind, magnetic field and rotation is likely responsible 
for the helium enrichment in the surface spots of early Bp stars. 
These stars exhibit strong ($B > 1$ kG) magnetic fields but weak stellar 
winds allowing substantial magnetic confinement (see ud-Doula \& Owocki, 
\cite{udD:Owo}).  Magnetic field strengths, however, 
should vary considerably from star to star, in particular if the fields are 
fossil. What would a star look like with only weak magnetic fields present 
(50-300 G) and a comparable wind as observed for $\sigma$\,Ori\,E ? 
Clearly in this case one would expect the wind to dominate over the magnetic 
field and align it radially at larger stellar distances as shown by MHD 
simulations of ud-Doula \& Owocki (\cite{udD:Owo}). As a consequence there 
are (nearly) no closed field lines, the wind can flow out freely unhindered 
by the field, no disk will be formed, and the magnetic field might escape 
detection.  
If the wind is not constant (we deal with decreasing wind at
decreasing metallicity), we expect the closing of more and more loops when the
wind becomes weaker, and therefore the eventual formation of a disk.

The most relevant physical parameters are then the relative strengths of wind
and magnetic field (i.e., ram pressure and magnetic pressure) and the value
$\beta$ of the magnetic inclination, i.e., the angle between rotation and 
magnetic field axis, assuming here that the large scale field is dipolar (which 
is not a necessary condition). Depending on the radiation field, fractionation 
will occur in the wind for helium (and possibly for hydrogen). Whenever 
particles of the passive elements do not reach escape velocity, they still 
have to follow open magnetic field lines, but once left to gravity they will 
fall back to the stellar surface, mainly along the same field lines. 
Hunger \& Groote (\cite{Hun:Gro}) therefore argued that 
fractionation is the source of helium enrichment in He-strong stars. 
The helium enrichment will increase the mass to be accelerated by the driving 
metals, whose abundance will on the other hand be reduced, leading to a decrease in 
wind velocity, thus increasing the reaccretion process, which in turn lowers 
the wind velocity even further.  Eventually magnetic field lines reconnect
and a disk is formed. On the other hand, once closed loops are present, 
helium particles will acquire sufficient velocity (although below escape 
velocity) to reach the disk.  Thus the process may be reversed and the 
helium enrichment is abolished. Alternatively, the helium layer on top of photosphere 
may become unstable, thus leading to mixing in the photosphere and
restoration of normal surface abundances. Once these are 
restored, the wind strengthens again, the matter in the disk is blown away, 
and magnetic reconnection is largely inhibited.  Thus the proposed scenario, 
although derived from Bp star modeling, may explain the hitherto little 
understood temporary appearance of circumstellar disks in Be-type stars, and 
thus {\em Be-type stars can be regarded as the weak field brothers of the 
Bp-type stars with small magnetic inclinations}.

Weak magnetic fields have already been proposed to explain the higher rotational
velocities of Be stars as compared to normal B stars. St\c{e}pie\'n (\cite{Ste}) 
attributes the larger rotational velocities of Be stars to spin up in their 
pre-main sequence phase and shows that magnetic fields above 400\,G lead to 
slow down, while small magnetic fields between 40 and 400\,G lead to an increase 
in rotational velocity. We then expect at least some helium enrichment also 
in Be stars, especially prior to the shell phase; such an He-enrichment 
of $N_{He}/N_{H} \approx 0.2$ was indeed determined for the Be star 60 Cygni 
(Koubsk\'y et al. \cite{Kub:al1}). 

The time interval between two shell phases should
strongly depend on the polar wind velocity.  If this scenario is correct,
X-ray flaring should be a common feature of both Bp and Be stars. Flares are 
expected to predominantly occur after magnetic field line reconnection but 
before or during shell formation. The flare in $\lambda$\,Eri might be 
interpreted as an indication of such an event, since two weeks after the 
flare there was no H$_{\alpha}$ emission observed (Smith et al., \cite{Smi:al1}).
At least at that time there had been no evidence for a 
disk around $\lambda$\,Eri.

\begin{acknowledgements}
We would like to thank M.A. Smith for a careful reading of our manuscript and 
extremely helpful comments concerning Be stars, as well as our referee
Dr. M. Gagn\'e whose criticisms helped to substantially improve our manuscript.

\end{acknowledgements}

\listofobjects

\begin{thebibliography}{}
\bibitem[1997a]{Bab:Mon} Babel, J., Montmerle, T. 1997a, \aap\ 323, 121
\bibitem[1997b]{Bab:Mo2} Babel, J., Montmerle, T. 1997b, \apj\ 485, L29
\bibitem[1994]{Ber:Sc1} Bergh"ofer, T.W., Schmitt, J.H.M.M. 1994, \aap\ 290, 435
\bibitem[1994]{Ber:Sc3} Bergh"ofer, T.W., Schmitt, J.H.M.M. 1994, \apss\ 221, 309
\bibitem[1994]{Ber:Sc4} Bergh"ofer, T.W., Schmitt, J.H.M.M. 1994, Science 
			265, 1689
\bibitem[1995]{Ber:Sc2} Bergh"ofer, T.W., Schmitt, J.H.M.M. 1995, Advances in 
			Space Research 16, 163
\bibitem[2002]{Cas:al1}	Cassinelli, J.P., Brown, J.C., Maheswaran, M.,
			Miller, N.A., Telfer, D.C. 2002, \apj\ 578, 951
\bibitem[1997]{Coh:al1}	Cohen, D.H., Cassinelli, J.P., MacFarlane, J.J. 1997, \apj\
			487, 867
\bibitem[2002]{Don:al1} Donati, J.-F., Babel, J., Harries, T.J. et al. 2002,
			\mnras\ 333, 55
\bibitem[1987]{Dra:al1} Drake, S.A., Abbot, D.C., Bastian, D.S. et al. 1987,
			\apj\ 322, 902
\bibitem[1997]{Gag:al1}	Gagn\'e, M., Caillault, J.-P., Stauffer, J.R.,
			Linsky, J.L. 1997, \apj 478, L78
\bibitem[2003]{Gro} 	Groote, D. 2003, in Magnetic fields in O, B
			and A stars: Origin and connection to pulsation,
			rotation and mass loss, Eds. L.A. Balona, H. Henrichs,
			T. Medupe, ASP Conference series, Vol.CS-305
\bibitem[1976]{Gro:Hu1} Groote, D., Hunger, K. 1976, \aap\ 52, 303
\bibitem[1977]{Gro:Hu2}	Groote, D., Hunger, K. 1977 \aap\ 56, 129
\bibitem[1982]{Gro:Hu3} Groote, D., Hunger, K. 1982, \aap\ 116, 64 (GH1)
\bibitem[1997]{Gro:Hu4} Groote, D., Hunger, K. 1997, \aap\ 319, 250
\bibitem[1984]{Hav:Goe} Havnes, O., Goertz, C.K. 1984, \aap\ 138, 421
\bibitem[1977]{Hes:al1} Hesser, J.E., Moreno, H., Ugarte, P. 1977, \apj\ 216, L31
\bibitem[1999]{Hun:Gro} Hunger, K., Groote, D. 1999, \aap\ 351, 554
\bibitem[1989]{Hun:al1} Hunger, K., Heber, U., Groote, D. 1989, \aap\ 224, 57
\bibitem[1990]{Hun:al2} Hunger, K., Heber, U., Groote, D. 1990, in Properties of 
			hot luminous stars, Proceedings of the First 
			Boulder-Munich Workshop, ed. Garmany, p. 307
\bibitem[1996]{Kue:Sch} K"urster, M., Schmitt, J.H.M.M. 1996, \aap\ 311, 211
\bibitem[2001]{Kri:Kub} Krti\v{c}ka, J., Kub\'at, J. 2001, \aap\ 369, 222
\bibitem[2000]{Kub:al1} Koubsk\'y, P., Harmanec, P., Hubert, A.M. et al. 2000,
			\aap\ 356, 913
\bibitem[1978]{Lan:Bor} Landstreet, J.D., Borra, E.F. 1978, \apj\ 224, L5
\bibitem[1985]{Lin} 	Linsky, J.L. 1985, Solar Physics 100, 333.
\bibitem[1982]{Luc} 	Lucy, L.B. 1982, \apj\  255, 286
\bibitem[2003]{Mah}	Maheswaran, M. 2003, \apj\ 592, 1156
\bibitem[1981]{Pal:al2} Pallavicini, R., Golub, L., Rosner, R., et al. 1981, 
			\apj\ 248, 279
\bibitem[2002]{Pal:al1} Pallavicini, R., Sanz-Forcada, J., Franciosini, E. 2002, 
			High Resolution X-ray Spectroscopy with XMM-Newton and 
			Chandra, Proceedings of the international workshop held 
			at the Mullard Space Science Laboratory of University 
			College London, Holmbury St Mary, Dorking, Surrey, UK, 
			October 24 - 25, 2002, Ed. G. Branduardi-Raymont,  
			published electronically and to be stored on CD., p. E29
\bibitem[1993]{Pat:Sim} Patten, B.M., Simon, T. 1993, \apj\ 415, L123
\bibitem[1999]{Por:Sk1} Porter, J.M., Skouza, B.A. 1999, \aap\ 344, 205
\bibitem[1989]{Phi:Les} Phillips, R.B., Lestrade, J.-F. 1988, Bull.A.A.S. 20, 728
\bibitem[1995]{Ran:al1} Randich, S., Schmitt, J.H.M.M., Prosser, C.F., Stauffer J.R. 1995,
			\aap\ 300, 134
\bibitem[2000]{Rei:al1} Reiners, A., Stahl, O., Wolf, B., Kaufer, A., Rivinius, T. 2000,
			\aap\ 363, 585
\bibitem[1999]{Sch:Fav} Schmitt, J.H.M.M., Favata, F. 1999, {\it Nature} 401, 44
\bibitem[1992]{Sch:al0} Schaller, G., Schaerer, D., Meynet, G., Maeder, A. 1992,
			\aaps\ 96, 269
\bibitem[1995]{Sch:al1} Schmitt, J.H.M.M., Fleming, T.A., Giampapa, M.S. 1995,
			\apj\ 450, 392
\bibitem[2003]{Sch:al3} Schmitt ,J.H.M.M., Ness, J.-U., Franco, G. 2003,
			\aap\ in press
\bibitem[1997]{Sch}  	Schmitt, J.H.M.M. 1997, A\&A 318, 215
\bibitem[2003]{Sch:Lie} Schmitt, J.H.M.M., Liefke, C. 2003, \aap\ in press
\bibitem[1990]{Sho:Bro} Shore, S.N., Brown, D.N. 1990, \apj\ 365, 665
\bibitem[1993]{Smi:al1} Smith, M.A., Grady, C.A., Peters, G.J., Feigelson, E.D. 1993, 
			\apj\ 409, L49
\bibitem[2001]{Smi:Gro} Smith, M.A., Groote, D. 2001, \aap\ 372, 20
\bibitem[2002]{Ste}	St\c{e}pie\'n, K. 2002, \aap\ 383, 21
\bibitem[2002]{udD:Owo} ud-Doula, A., Owocki, S.P. 2002, \apj\ 576,413
\end{thebibliography}
\end{document}